\documentclass[aps,prb,a4paper,preprint,showpacs]{revtex4}
\usepackage{graphicx}
\usepackage{amsmath}
\usepackage{amsfonts}
\usepackage{amssymb}
\usepackage{slashbox}
\usepackage{amsbsy}
\usepackage{bm}
\usepackage{epstopdf}
\usepackage{epsfig}

\begin{document}
\title{ Fano resonance via bonding and antibonding states in nonadiabatically-pumped double-quantum-well systems }
\author{Cong Cai and Rui Zhu\renewcommand{\thefootnote}{*}\footnote{Corresponding author.
Electronic address:
rzhu@scut.edu.cn}}
\affiliation{ Department of Physics, South China University of Technology,
Guangzhou 510641, People's Republic of China}

\begin{abstract}

 In this work, transport properties of the nonadiabatically pumped double-quantum-well (DQW) structure are studied. Different from a single quantum well, band mixing in the DQW generates bonding and antibonding states, whose wave functions have different spatial symmetry. By applying a time-dependent electric potential to the two well regions simultaneously, Floquet sidebands are formed, which constitutes additional quantum tunneling paths. When one of the Floquet sidebands coincides with the bonding or antibonding quasibound states within the DQW structure, sharp Fano resonances are found in the transmission coefficients as well as in the differential shot noise spectra. While such Fano resonances originate from quantum interference, their shapes are strikingly different for transport via the bonding state and via the antibonding state. The Fano resonance via the even-parity bonding state shows a perfect transmission followed by a total reflection and the Fano resonance via the odd-parity antibonding state has a reversed symmetry and shows a total reflection before a perfect transmission.

\end{abstract}

\pacs {73.21.Fg, 72.70.+m, 72.30.+q}

\maketitle
\section{Introduction}

 Generally the concept of Fano resonance roots in quantum interference and asymmetric resonant line profile, the latter of which originates from a close coexistence of resonant transmission and resonant reflection and can be interpreted by the interaction of a discrete localized state with a continuum of propagation modes\cite{42MiroshnichenkoRMP2010, 43FanoPR1961, TekmanPRB1993}. It is a
universal effect existing almost in all interfering quantum
processes, independent of the specific details of the system
under study. The interfering paths can be formed by
spatial inhomogeneity as well as time-dependent oscillation.
Subsequent to its discovery, there have been a great
number of studies devoted to Fano resonances in various
quantum systems, such as Anderson impurity systems\cite{HGLuoPRL2004},
quantum dots\cite{JohnsonPRL2004, BulkaPRL2001, TorioEPJB2004}, scattering from a donor impurity in an
electron waveguide\cite{TekmanPRB1993, CSChuPRB1989, RuiZhuJPCM2013}, tunneling through an ${\rm{Al}}_{\rm{x}} {\rm{Ga}}_{{\rm{1 - x}}} {\rm{As}}$ barrier\cite{TekmanPRB1993, BoykinPRB1992}, transmission through a waveguide linked to a resonant cavity\cite{TekmanPRB1993, PorodAPL1992}, spin inversion devices\cite{CardosoEPL2008}, nanowires
and tunnel junctions\cite{KobayashiPRB2004}, plasmonic nanoparticles,
photonic crystals, electro-magnetic metamaterials\cite{LukyanchukNatMat2010}, and etc. The
steep dispersion of the Fano resonance profile promises applications in sensors, lasing, switching, and nonlinear and
slow-light devices\cite{42MiroshnichenkoRMP2010, LukyanchukNatMat2010}.
In the present approach, a specific mechanism to generate Fano resonance is considered--Floquet scattering in nonadiabatic quantum pumping. Physics of the nonadiabatic quantum pumping is different from its adiabatic counterpart described by the Berry phase of the scattering matrix enclosed by the cyclic trajectory in the parameter space\cite{1ThoulessPRB1983, 4XiaoRMP2010, 3BrouwerPRB1998, 2MoskaletsPRB2002}. When a high-frequency ac potential is applied to a tunneling junction, the Floquet sidebands are formed, which supplies additional quantum paths for interference to occur. It has been predicted that in the ac-potential-driven quantum well, Fano resonance can be observed in the transmission coefficients, Wigner-Smith delay times, conductance and shot noise spectrum when one of the Floquet sidebands coincides with one of the quasibound states confined in the well\cite{38LiPRB1999, 44Jiao-Hua DaiEPJB2014}.

 Although recently some attention has been paid to Floquet scattering in nonadiabatically pumped tunneling systems\cite{EckardtRMP2017, WeiYinDengNJP2015, UsajPRB2014, DelplacePRB2013, LopezPRB2012, 38LiPRB1999, 16S.L.ZhuPRB2002, 44Jiao-Hua DaiEPJB2014, 49RZhuJAP2015, 26Y.-C. XiaoPLA2013, 36San-JosePRB2011, RuiZhuEPJB2016January}, such as topological insulator\cite{WeiYinDengNJP2015}, graphene\cite{36San-JosePRB2011, UsajPRB2014, DelplacePRB2013, LopezPRB2012}, superconductor Josephson junction\cite{RuiZhuEPJB2016January}, and etc., Fano resonance was not observed in all of the materials or devices. In graphene and two-dimensional electron gas (2DEG), Fano resonance occurs when the
wavevector in the transport direction of one of the Floquet sidebands is exactly identical to that of
the quasibound state confined in a quantum well. Using the resonance position in the pumped shot noise spectrum, one can exactly trace the dispersion pattern of the quasibound state\cite{49RZhuJAP2015}. And the nonadiabatic transport properties of the one-dimensional time-dependent superconductor Josephson junction display resonances when one of the electron or hole Floquet wavevectors
coincides with the Andreev bound states within the superconducting energy gap. The resonance
varies with the phase difference between the two superconductors as a result of the bound-level displacement\cite{RuiZhuEPJB2016January}. These observations prompt us to pursue the physical mechanism of nonadiabatically pumped Fano resonance to a more fundamental level, i.e., relation between properties of the multiple quantum paths involved and the Fano resonance spectrum.

Towards this target, we consider the double-quantum-well (DQW) tunneling junction driven by a time-dependent potential applied to the two well regions simultaneously. The advantage of this device is that the bonding and antibonding states are formed by wavefunction overlapping of the two wells. The real part of bonding wavefunction has an even parity while that of the antibonding wavefunction has an odd parity. The proposed device is of practical meaning in semiconductor heterostructures\cite{52S. FafardPRB1993, 57KrausePRB1998, 55BetancourtRieraPhysica E2012}, laser-induced photonic superstructure\cite{53S. M. SadeghiPRB2010}, coupled quantum dots\cite{54SchedelbeckScience1997}, and etc. Using the Floquet scattering theory, transport properties of the dynamic tunneling process are investigated. As a result of spatial- and time-reversal symmetry, no net dc current can be pumped out from one lead to the other, therefore, transmission coefficients and pumped shot noise spectrum are numerically obtained. Electrical shot noise is defined by quantum correlation of the current operator and measures time fluctuations of the current originating from quantization of charge carriers\cite{45BlanterPR2000}. This definition of shot noise is shared among the static transport\cite{45BlanterPR2000}, adiabatic quantum pumping\cite{2MoskaletsPRB2002}, and nonadiabatic quantum pumping\cite{6MoskaletsPRB2004} with its physical mechanism slightly different. When the conductance is vanishingly small, the shot noise and Fano factor defined by noise-to-current ratio can reflect the transmission properties as well. This has been very useful in graphene\cite{TworzydloPRL2006} and pseudospin-1 Dirac-Weyl systems\cite{ZhuHuiPLA2017}. In nonadiabatic quantum pumping, correlation among different Floquet channels secures a significant shot noise even when time- and spatial-reversal symmetry excludes a net dc current flow\cite{6MoskaletsPRB2004}. Therefore, relation between the wavefunction symmetry and nonadiabatically pumped shot noise is of the main focus in our final results.

\section{Model and formalism}

We consider a one-dimensional time-dependent double potential-well sketched in Fig. 1. The double-quantum-well (DQW) structure is formed with two identical potential wells with $L$ the width and $U_0$ the depth of each of them separated by an zero-energy interval of length $b$. A time-dependent external electric gate potential of frequency $\omega$ and strength $U_1$ is applied to the two well regions simultaneously. Such a physical situation can be described by the time-dependent Schr\"{o}dinger equation
\begin{equation}
i\hbar \frac{\partial }{{\partial t}}\Psi \left( {x,t} \right) = \hat H\left( {x,t} \right)\psi \left( {x,t} \right),
\label{SchrodingerEquation}
\end{equation}
with
\begin{equation}
\hat H(x,t)=\frac{\hbar^{2}}{2m^{\star}}\frac{\partial^{2}}{\partial{x^{2}}}+U(x,t),
 \end{equation}
and
\begin{align}
U(x,t)=\begin{cases}
-U_0+U_1\cos(\omega t),& 0<x<L, L+b<x<2L+b,
\\
0,&\text{others}.
\end{cases}
\end{align}
Here, $m^{\star}$ is the effective mass of electrons.

 Before going on to the time-dependent problem, it is meaningful to consider the quasibound states within the static DQW structure first. When the electron energy $E$ lies in the regime with $-U_0<E<0$, the wave function can be expressed as
\begin{align}
\begin{cases}
re^{Kx},&x<0,\\
ae^{-ikx}+be^{ikx},&0<x<L,\\
ce^{Kx}+de^{-Kx},&L<x<L+b,\\
fe^{-ikx}+ge^{ikx},&L+b<x<2L+b,\\
te^{-Kx},&x>2L+b,
\end{cases}
\end{align}
where $K=\sqrt{2m^{\star}(-E)}/\hbar$ and $k=\sqrt{2m^{\star}(E+U_0)}/\hbar$. Continuity of the wave function and its first-order derivative gives rise to the secular equation
\begin{equation}
\left|\begin{array}{cccccccc}
1&-1&-1&0&0&0&0&0\\
K&ik&-ik&0&0&0&0&0\\
0&e^{-ikL}&e^{ikL}&-e^{KL}&-e^{-KL}&0&0&0\\
0&-ike^{-ikL}&ike^{ikL}&-Ke^{KL}&Ke^{-KL}&0&0&0\\
0&0&0&e^{K(L+b)}&e^{-K(L+b)}&-e^{-ik(L+b)}&-e^{ik(L+b)}&0\\
0&0&0&Ke^{K(L+b)}&-Ke^{-K(L+b)}&ike^{-ik(L+b)}&-ike^{ik(L+b)}&0\\
0&0&0&0&0&e^{-ik(2L+b)}&e^{ik(2L+b)}&-e^{-K(2L+b)}\\
0&0&0&0&0&-ike^{-ik(2L+b)}&ike^{ik(2L+b)}&Ke^{-K(2L+b)}
\end{array}\right|=0,
\end{equation}
from which the eigenenergies of the confined states can be obtained. By substituting these eigenenergies into the continuity equations, the corresponding wave function can be obtained. Properties of the quasibound states will be discussed in the next section.

In order to investigate transport properties of the time-dependent DQW with a constant varying frequency, we use the Floquet scattering theory and consider the following region-wise wave function
\begin{equation}
\begin{aligned}
\Psi_1(x,t)&=\sum\limits_{n=-\infty}^{\infty}e^{-iE_{n}t/\hbar}(a_{n}^{l}e^{ik_{n}x}+b_{n}^{l}e^{-ik_{n}x}),x\le0 , \\
\Psi_2(x,t)&=\sum\limits_{n=-\infty}^{\infty}e^{-iE_{n}t/\hbar}\sum\limits_{m=-\infty}^{\infty}(a_{m}e^{-iK_{m}x}+b_{m}e^{iK_{m}x})\times{J_{n-m}(\frac{U_1}{\hbar\omega})},
0\le{x}\le{L} , \\
\Psi_3(x,t)&=\sum\limits_{n=-\infty}^{\infty}e^{-iE_{n}t/\hbar}(c_{n}e^{-ik_{n}x}+d_{n}e^{ik_{n}x}),L\le{x}\le{L+b} , \\
\Psi_4(x,t)&=\sum\limits_{n=-\infty}^{\infty}e^{-iE_{n}t/\hbar}\sum\limits_{m=-\infty}^{\infty}(c_{m}e^{-iK_{m}x}+d_{m}e^{iK_{m}x})\times{J_{n-m}(\frac{U_1}{\hbar\omega})},
L+b\le{x}\le{2L+b}, \\
\Psi_5(x,t)&=\sum\limits_{n=-\infty}^{\infty}e^{-iE_{n}t/\hbar}(a_{n}^{r}e^{-ik_{n}x}+b_{n}^{r}e^{ik_{n}x}),x\ge2L+b , \\
\end{aligned}
\end{equation}
with ${\Psi _i}\left( {x,t} \right)$ ($i = 1,2, \cdots ,5$) denoting $\Psi \left( {x,t} \right)$ in the $i$-\emph{th} region. Here, $a_{n}^{l/r}$ and $b_{n}^{l/r}$ are the probability amplitudes corresponding to the incoming and outgoing electron waves of the left/right lead, respectively. $E_{n}=E_{F}+n\hbar\omega$ is the eigenvalue of the $n$-\emph{th} order Floquet state, where $E_{F}$ is the Fermi energy and $k_n=\sqrt{2m^{\star}E_n}/\hbar$. In the exact case $n$ are integers ranging from $-\infty$ to $\infty$. In numerical considerations, we use the justified cutoff\cite{38LiPRB1999} of $\left| n \right| \le N$ with $N =5 > {{{U_1}} \mathord{\left/
 {\vphantom {{{U_1}} {\hbar \omega }}} \right.
 \kern-\nulldelimiterspace} {\hbar \omega }}$. For $E_{n}<0$, i.e., when $k_{n}$ is imaginary, the corresponding Floquet mode is evanescent, which does not directly contribute to the pumped current and noise. $K_{m}=\sqrt{2m^{\star}(E_{F}+m\hbar\omega)}/\hbar$ is the wavevector in the oscillating quantum well region. $J_n (x)$ is Bessel function of the first kind originating from separation of variables towards the time-dependent Schor\"{o}dinger equation (\ref{SchrodingerEquation}) by $\exp \left[ { - {\textstyle{i \over \hbar }}\int_0^t {{U_1}\cos \left( {\omega t'} \right)dt'} } \right] = \sum\limits_{n =  - \infty }^\infty  {{J_n}\left( {{\textstyle{{{U_1}} \over {\hbar \omega }}}} \right){e^{ - in\omega t}}} $. $c_n$ and $d_n$ are the probability amplitudes in the middle region can be obtained by considering the continuity relation at the four boundaries together with $a_{n}^{l/r}$ and $b_{n}^{l/r}$. $a_m$, $b_m$, $c_m$, and $d_m$ constitute matrices indexed by row number $n$ and column number $m$ in the $2$-\emph{nd} and $4$-\emph{th} region and can also be solved implicitly by the continuity equations. These continuity equations are that the wave function and its first-order spatial derivative is continuous at interfaces.

With all the probability amplitudes obtained, we can connect the incoming and outgoing modes outside of the DQW by the matrix $S$ as
\begin{equation}
\left(\begin{matrix}
b_{n}^{l}\\
b_{n}^{r}
\end{matrix}\right)=\sum\limits_{m}\left(\begin{matrix}
r_{nm}&t ' _{nm}\\
t_{nm}&r ' _{nm}
\end{matrix}\right)\left(\begin{matrix}
a_{m}^{l}\\
a_{m}^{r}
\end{matrix}\right)\\
=\sum\limits_{m}S_{nm}\left(\begin{matrix}
a_{m}^{l}\\
a_{m}^{r}
\end{matrix}\right).
\end{equation}
The elements in the Floquet scattering matrix $r_{nm}$/$r ' _{nm}$ and $t_{nm}$/$t ' _{nm}$ are the reflection and transmission amplitudes, respectively, incident from the $m$-\emph{th} Floquet channel and scattered into the $n$-\emph{th} channel. The non-primed correspond to those outgoing from the left reservoir while the primed correspond to those outgoing from the right. To consider the real current flux, we express the Floquet scattering matrix $s$ as
\begin{equation}
s\left( {{E_n},{E_m}} \right) = \left\{ {\begin{array}{*{20}{l}}
{\sqrt {\frac{{{\mathop{\rm Re}\nolimits} {k_n}}}{{{\mathop{\rm Re}\nolimits} {k_m}}}} {S_{nm}},}&{{\mathop{\rm Re}\nolimits} {k_m} \ne 0,}\\
{0,}&{{\mathop{\rm Re}\nolimits} {k_m} = 0.}
\end{array}} \right.
\label{FloquetMatrix}
\end{equation}
From Eq. (\ref{FloquetMatrix}) we can see that the scattering probabilities originating from both the incoming and outgoing evanescent modes are zero and therefore such channels do not contribute to the transport process. With the scattering matrix obtained, we can define the total transmission probability as
\begin{equation}
T = \sum\limits_{m =  - \infty }^\infty  {{{\left| {s\left( {{E_0},{E_m}} \right)} \right|}^2}} .
\label{TotalTransmission}
\end{equation}

In our present consideration, the two quantum wells are identical in depth and width and are subject to the same time-dependent gate potential as shown in Fig. 1. In this case, both the time- and spatial- reversal symmetries are strictly preserved giving rise to a vanishing pumped charge current. We set the parameters in this way to most prominently demonstrate the relation between the shapes of the Fano resonance and the bound-state wave function, discussions of which will be given in the next section.

While the pumped charge current does not exist, we go on to its first-order correlation function--the shot noise. The pumped shot noise correlating the $\alpha$ and $\beta$ leads is defined as\cite{45BlanterPR2000,58ButtikerPRB1992}
\begin{equation}
S_{\alpha\beta}(t_1,t_2)=\frac{1}{2}\langle\hat{I}_\alpha(t_1)\hat{I}_\beta(t_2)+\hat{I}_\beta(t_2)\hat{I}_\alpha(t_1)\rangle
\label{DefiningShotNoise}
\end{equation}
where $\hat{I}_\alpha(t)$ is the quantum-mechanical current operator in the lead $\alpha$ and
\begin{equation}
\hat{I}_\alpha(t)=\frac{e}{h}\int{dE}d{E^{'}}[\hat{b}_{\alpha}^{\dagger}(E)\hat{b}_{\alpha}(E^{'})-\hat{a}_{\alpha}^{\dagger}(E)\hat{a}_{\alpha}(E^{'})]e^{i(E-E^{'})t/\hbar}.
\end{equation}
The annihilation operators $\hat{a}_{\alpha}(E)$ and $\hat{b}_{\alpha}(E)$ apply to the incident and outgoing electrons, respectively. They are connected by the scattering matrix as
\begin{equation}
\hat{b}_{\alpha}(E)=\sum\limits_{n,\beta}s_{\alpha\beta}(E,E_n)\hat{a}_\beta(E_n).
\label{SMatrixConnectingOperators}
\end{equation}
From equations (\ref{DefiningShotNoise}) to (\ref{SMatrixConnectingOperators}), we can obtain the  pump shot noise\cite{6MoskaletsPRB2004}
\begin{equation}
S_{\alpha\beta}=\frac{e^{2}}{h}\int_{0}^{\infty} dE \sum\limits_{\gamma,\delta}\sum\limits_{n,m,p=-\infty}^{\infty}M_{\alpha\beta\gamma\delta}(E,E_m,E_n,E_p)[f_0(E_n)-f_0(E_m)]^{2},
\label{ShotNoise}
\end{equation}
with
\begin{equation}
M_{\alpha\beta\gamma\delta}(E,E_m,E_n,E_p)=s_{\alpha\gamma}^{\star}(E,E_n)s_{\alpha\delta}(E,E_m){s_{\beta\delta}^{\star}(E_p,E_m)s_{\beta\gamma}(E_p,E_n)}.
\end{equation}
As a result of current flux conservation, we have $S_{LL}=S_{RR}=-S_{LR}=-S_{RL}$. Therefore, we only consider one of them and label $S_{LL}$ as $S_I$. Because the shot noise measured in Eq. (\ref{ShotNoise}) is an accumulating result of all propagating Floquet channels around the Fermi energy, it is convenient to consider the differential shot noise defined as follows to demonstrate the Fano resonance occurring sharply at a certain energy more prominently
\begin{equation}
S_{I}^{d}=\frac{d{S_I}}{d {E_F}}.
\label{DifferentialShotNoise}
\end{equation}

\section{Results and discussions}

Firstly we present properties of the bonding and antibonding quasibound states confined in the DQW system. Energies of the quasibound states as a function of $L$ with fixed $U_0$ and $b$ are shown in Fig. 2 and wavefunctions of a pair of bonding and antibonding states are shown in Fig. 3. From fundamental quantum mechanics, we know that the bound-state energy of a single finite-depth quantum well is inversely proportional to the well width. And in the case of a DQW, coupling between the two wells doubles each of the single-well bound states giving rise to a bonding state with lower energy and antibonding state with higher energy. We can see from Fig. 2 that $E_b$ is approximately proportional to $1/L^2$ with the deviation coming from the coupling between the two wells. Numerical results also demonstrate that $E_{b1}<E_b$ and $E_{b2}>E_b$ with $E_b$ the same single-quantum-well bound-state energy, which means $E_{b1}$ is the corresponding bonding state and $E_{b2}$ the antibonding state. To further demonstrate this point, we plot the wavefunctions of the two states in Fig. 3. Wavefunctions of the bonding and antibonding quasi-bound states have significant difference. We can observe from Fig. 3 that the bonding state combines the two wells to each other, and the antibonding state separates them from each other. This property of the bonding and bonding states can be clearly seen in the probability density distribution shown in Fig. 3 (c). The maximum point of the bonding state moves closer towards the middle and that of the antibonding state moves farther away from the middle. The probability density of the antibonding state vanishes in the middle of the two wells while that of the bonding state sustains a considerably large value. The other significant difference between the bonding and bonding states is in the spatial symmetry. It can be seen in Fig. 3 (a) that the real part of the bonding wavefunction has an even parity while that of the antibonding wavefunction has an odd parity. The consequence of this difference to the Fano resonance profile will be shown now.

Numerical results of the total transmission probability obtained by Eqs. (\ref{TotalTransmission}) are shown in Fig. 4 (a). Fano resonance can be observed in the transmission probability spectrum at $E_{\rm{Fano1}} \approx 3.5943$ meV and $E_{\rm{Fano2}} \approx 10.7262$ mev for the case of $\hbar\omega=14$ mev (referred to the blue solid curve in Fig.4 (a)). Relationship of the quasi-bound energy and resonance energy is that $E_{\rm{Fano1/Fano2}}=E_{b1\rm{/}b2}+\hbar\omega$. When an electron incident from the left reservoir with energy $E_F$ loses a photon of energy $\hbar\omega$, drops to the ``bound" state, at the same time absorbs a photon of energy $\hbar\omega$, and jumps back to the incident channel towards the right reservoir, Fano resonance occurs as a result of interference between direct tunneling and tunneling via the bound level through Floquet scattering. It should be noted that moving of the $\hbar\omega$ energy quanta is sustained by the external driving force, who drives the energy flow. A remarkable feature in the numerical results is that the two resonant profiles have different spatial symmetry. The Fano resonance originating from the bonding state has a resonant maximum followed by an antiresonant minimum. Fano resonance originating from the antibonding state has an antiresonant minimum followed by a resonant maximum. We attribute this difference to the parity reversal in the real part wavefunctions of the bonding and antibonding states as the Fano resonance is a direct result of quantum interference via the two kinds of quantum levels. Apart from the resonance-antiresonance reversal, shapes of the two kinds of Fano resonances have another difference. The range of $E_{\rm{Fano2}}$ is wider than that of $E_{\rm{Fano1}}$. This also originates from the difference in the wavefunctions of the bonding and antibonding states. In the bonding state, states of the two wells tend to attract each other and the wavefunction contracts into a smaller range. In the antibonding state, states of the two wells tend to repel each other and the wavefunction extends into a larger range. Therefore, the shape of Fano resonances follows the shape of wavefunctions generating the resonances.

Fig. 4 (b) and (c) show results of the shot noise and differential shot noise as a function of the Fermi energy obtained by Eqs. (\ref{ShotNoise}) and (\ref{DifferentialShotNoise}), respectively. With the increase of the Fermi energy, noise is increased while more energy channels are below the Fermi energy and contribute to the transport. Because all channels below the Fermi energy contribute to the shot noise and the Fano resonance occurs at a single energy in the transmission probability, the shot noise spectrum only demonstrates an inflection at the Fermi energy where the Fano resonance appears in the transmission spectrum. By differentiating the shot noise with respect to the Fermi energy as formulated in Eq. (\ref{DifferentialShotNoise}), the effect occurring in a single energy can be magnified and the Fano resonance becomes obvious in the differential shot noise shown in Fig. 4 (c). By further inspection of Fig. 4 (a) and (c), a significant difference between them cannot evade a careful observer. Wherever the transmission demonstrates a resonant maximum of perfect transmission, the differential shot noise demonstrates a minimum; and wherever the transmission demonstrates a sharp drop to the resonant minimum of total reflection, the differential shot noise demonstrates a maximum. This behavior can be satisfactorily interpreted by general theories of the shot noise\cite{45BlanterPR2000}. For a single-channel transmission the shot noise approximates a value of $T(1-T)$ with $T$ the transmission probability. This relation demonstrates that when $T=1$ and $0$ the shot noise vanishes and when $T=0.5$ the shot noise maximizes. Therefore it is no wonder that the differential shot noise reflecting transport of a single energy channel vanishes at prefect transmission and maximizes in the middle of the sharp drop from $T=1$ to $T=0$. No matter the difference between the differential noise and the transmission, the spatial-symmetry reversal between $E_{\rm{Fano1}}$ and $E_{\rm{Fano2}}$ is inherited in the former from the latter. The wavefunction difference thus becomes experimentally expressible by the Fano resonance in  the shot noise.

Besides the numerical results shown in the figures, we also applied other parameter settings to confirm that the behavior of the nonadiabatically pumped shot noise is similar in all shapes of DQW structures.
We considered the case of smaller or larger well separations and well widths and found that spatial-symmetry reversal in the Fano resonance in the nonadiabatic transmission probability and differential shot noise maintains as a result of the parity reversal between the bonding and antibonding wavefunctions. We also considered the case of asymmetric DQW structures with widths of the two wells different and found similar results. In all the cases the Fano resonance occurs at the Fermi energy $E_{\rm{Fano}}=E_b+ \hbar \omega$ and the shape of its profile becomes wider or thinner following that of the quasibound state. In configurations without a temporal- and spatial-reversal symmetry such as the asymmetric DQW system, Fano resonance in the transmission probability is demonstrated in the differential pumped current as well as the shot noise.

\section{Conclusions}

In this work, we used the Floquet scattering theory to study the transport properties of the time-dependent DQW systems. Floquet transmission probabilities, shot noise, and the differential shot noise as a function of the Fermi energy are obtained. We discovered that two Fano resonances are excited while one of the Floquet sidebands coincides with the bonding and antibonding quasibound states, respectively. Because Fano resonance is the result of quantum interference between direct tunneling and Floquet tunneling via the quasibound state, the two Fano resonances demonstrate complete different spatial symmetry following the reversed parity of the bonding and antibonding wavefunctions. We also discussed the robustness of such a relation by numerical results of DQW structures of various shapes. In configurations with a temporal- and spatial-reversal symmetry, Fano resonance in the transmission probability is demonstrated in the differential shot noise. In configurations without a temporal- and spatial-reversal symmetry, Fano resonance in the transmission probability is demonstrated in the differential pumped current as well as the shot noise. In conclusion, symmetry in profiles of the wavefunctions of static quasi-bound states determines that of the Fano resonance in the quantum transport quantities.

We close with another prospect of the present work. While in all the nonadiabatically pumped systems hosting discrete localized states ever studied, Fano resonance is observed in the transmission spectrum when one of the Floquet sidebands coincides with the discrete level\cite{RuiZhuEPJB2016January, 49RZhuJAP2015}, and particularly in this work, dependence of the resonant profile on wavefunction parity is highlighted, it is not ungrounded to predict that Fano resonance can be used to detect and characterize properties of other kinds of quasibound states and discrete levels. Among these, we would like to mention some relevant to the contemporary physical society: 1. Majorana bound states, which is special kinds of Andreev bound states hosted in topological superconductors\cite{LutchynPRL2010, LiangFuPRL2008}; 2. Landau levels in magnetic-field-penetrating 2DEG, in which single-quantum-well or DQW system can be built in multilayer heterostructures\cite{56Song HePRB1991}; 3. Landau levels in skyrmion crystals under emergent electrodynamics\cite{HamamotoPRB2015}. Although mathematical formalisms of the latter two are remarkably different from the reference works, general theory of the Fano resonance--interaction of a discrete (localized) state with a continuum of propagation modes, should also see success in these systems.

\section{Acknowledgements}

This project was supported by the National Natural Science
Foundation of China (No. 11004063) and the Fundamental Research
Funds for the Central Universities, SCUT (No. 2017ZD099).

\clearpage

\begin{figure}[ht]
\includegraphics[height=12cm, width=10cm]{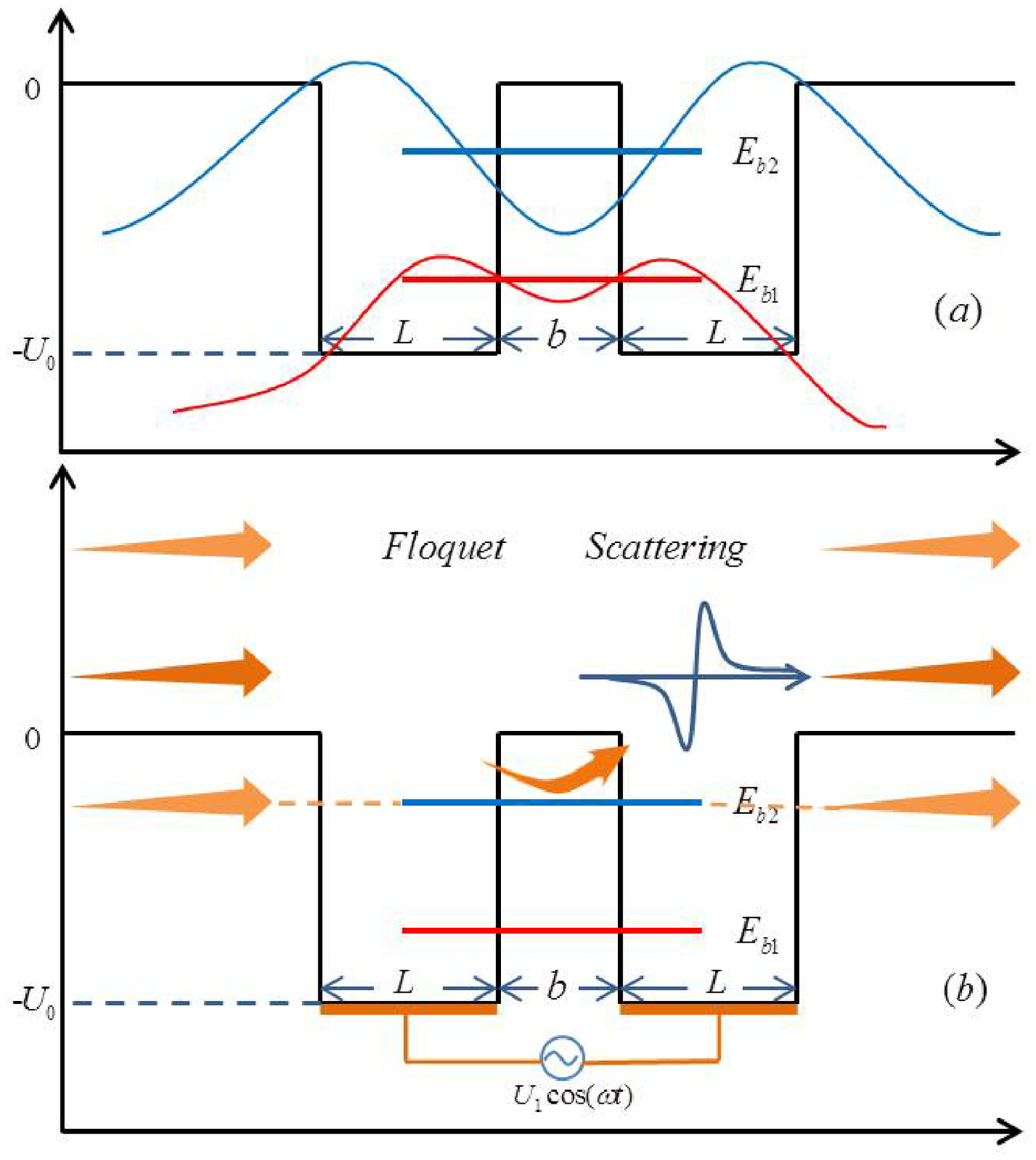}
\caption{ Schematic illustration of the non-adiabatically pumped DQW. Panel (a) sketches the static DQW without a driving potential. The two wells are identical in width and depth separated by a ground-level potential barrier. Confined in the DQW are two quasi-bound states, i.e., the bonding $E_{b1}$ and anti-bonding state $E_{b2}$ originating from the same single-well bound state $E_b$. The blue and red curves are corresponding probability densities of the bound states. In the antibonding state, the two wells are decoupled with the probability density touching zero in the middle, while in the bonding state, the two wells are coupled with significantly large probability density in the middle. Panel (b) sketches the Floquet scattering process. Floquet modes serve as additional quantum paths in tunneling. Especially when one of the Floquet sidebands coincides with the bound state, interference between paths of direct tunneling and sideband tunneling via the bound state gives rise to a strong Fano resonance in the transmission spectrum. The equilibrium well depth of the DQW is $U_0$. Width of the two wells is $L$ and their distance is $b$. The ac driving voltage is applied to the two wells simultaneously with an amplitude of $U_1$.
 }
\end{figure}

\begin{figure}[ht]
\includegraphics[height=10cm, width=12cm]{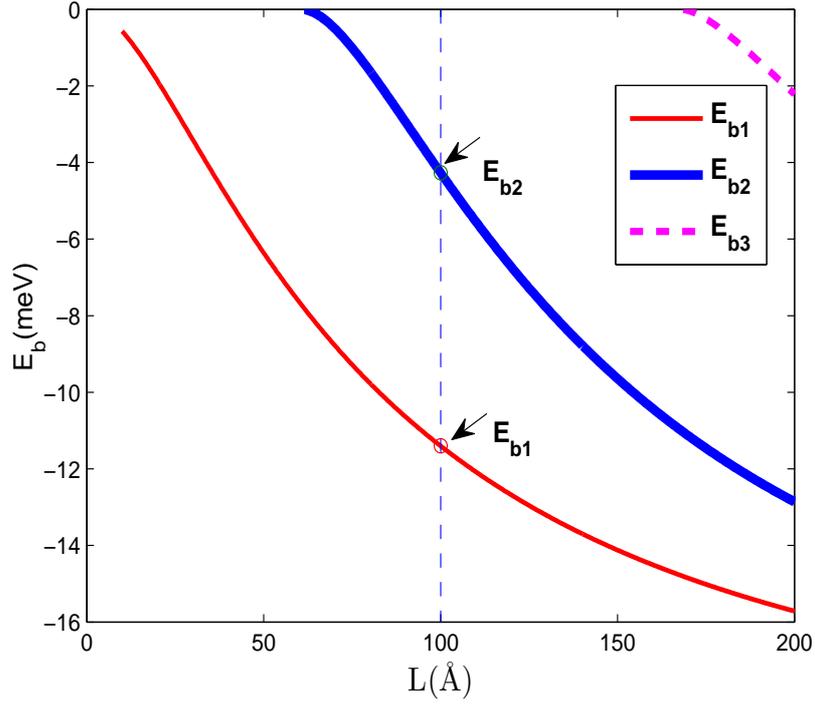}
\caption{ Energies of the quasibound states as a function of the well width $L$ with $U_0=20$ meV and $b= 50$ {\AA}. Our main numerical results consider the case of $L=100$ \AA with $E_{b1}\approx-11.4057$ meV and $E_{b2}\approx-4.2738$ meV indicated by a vertical blue dashed line and two circles. }
\end{figure}

\begin{figure}[ht]
\includegraphics[height=10cm, width=12cm]{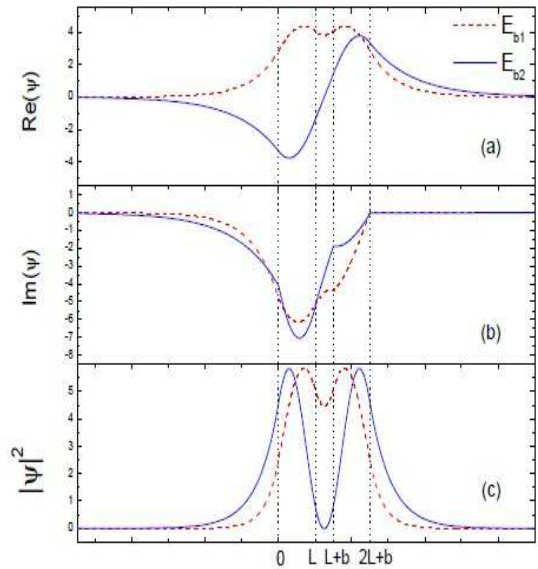}
\caption{Real part (a), imaginary part (b), and probability density (c) of the wavefunctions of the bonding and antibonding states. Dashed and solid lines describe the bonding and antibonding wavesfunctions, respectively. It can be seen in panel (a) that real part of the bonding state has an even parity and that of the antibonding state has an odd parity. It can be seen in panel (c) that the two wells repel each other in the antibonding state giving rise to a vanishing probability in the middle and the two wells attract each other in the bonding state giving rise to a considerably large probability in the middle. Also as a result of wavefunction attracting or repelling, maximum of the probability density occurs closer to the middle in the bonding state and farther away from the middle in the antibonding state. In both cases, the probability distribution has a spatial-reversal symmetry. }
\end{figure}

\begin{figure}[ht]
\includegraphics[height=12cm, width=14cm]{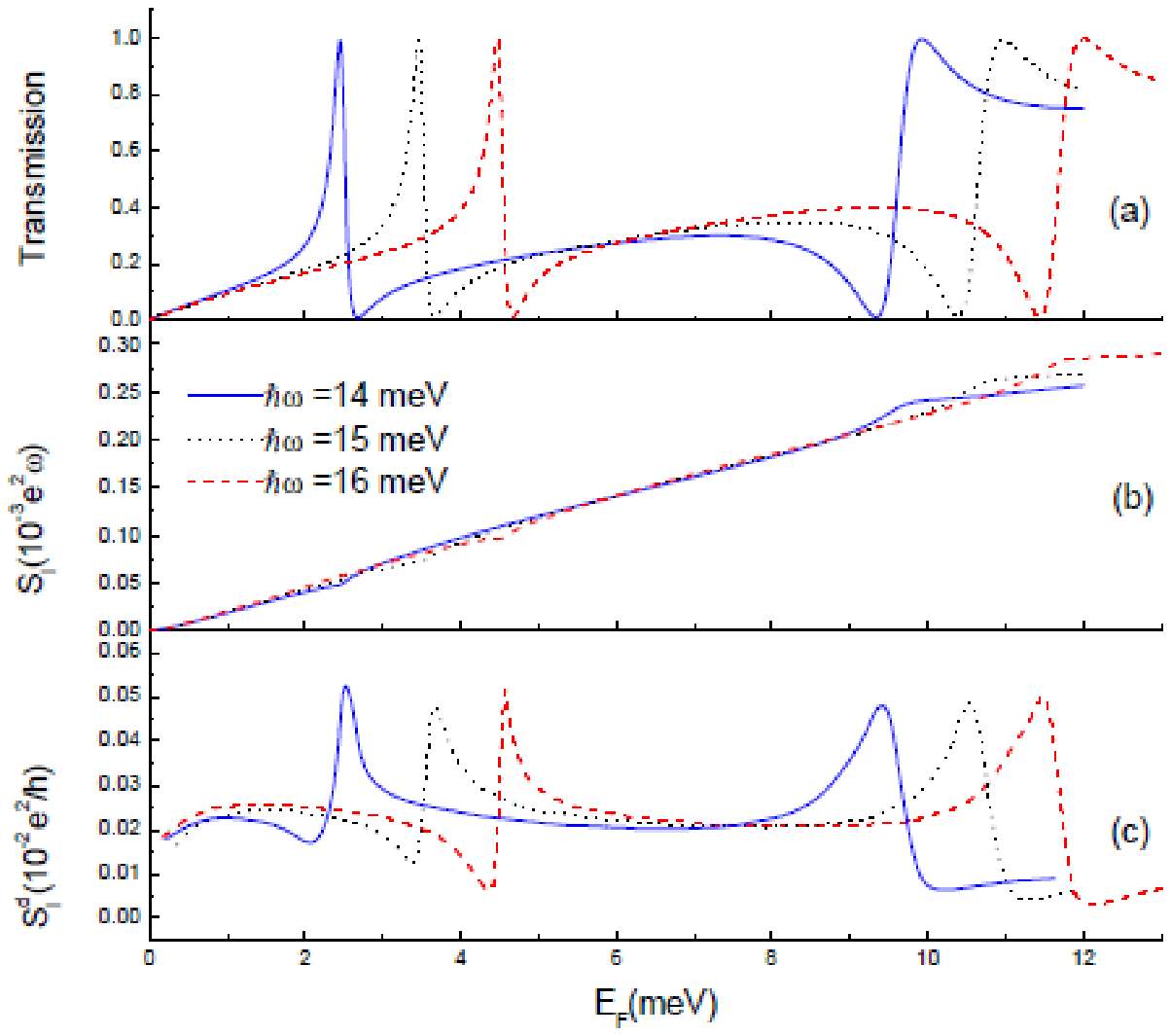}
\caption{(a) Total transmission probability $T$, (b) shot noise $S_{I}$, and (c) differential shot noise $S_{I}^{d}$ as a function of the Fermi energy in the nonadiabatically pumped DQW system. The three curves correspond to different driving frequencies in all panels. Sharp Fano resonances occur twice at $E_{F}=E_{b 1{\rm{/}}b2}+ \hbar\omega$ in the total transmission and differential shot noise for each variation along the Fermi energy, which originates from Floquet scattering via the bonding state $E_{b1}$ and antibonding state $E_{b2}$, respectively. Other parameters numerically used are the driving amplitude $U_1=8$ meV, static well depth $U_0=20$ meV, well width $ L=100 \rm{\mathring{A}}$ and separation $b= 50 \rm{\mathring{A}}$. Unit in the figures is obtained by substituting $\hbar\omega$ into the energy and absorbing additional $2 \pi$ into the data.}
\end{figure}


\clearpage
\begin{references}

\bibitem{42MiroshnichenkoRMP2010} A. E. Miroshnichenko, S. Flach, and Y. S. Kivshar, Rev. Mod. Phys.\textbf{82}, 2257 (2010).

\bibitem{43FanoPR1961} U. Fano, Phys. Rev. \textbf{124}, B1886 (1961).

\bibitem{TekmanPRB1993} E. Tekman and P. F. Bagwell, Phys. Rev. B \textbf{48}, 2553 (1993).

\bibitem{HGLuoPRL2004} H. G. Luo, T. Xiang, X. Q. Wang, Z. B. Su, and L. Yu, Phys. Rev. Lett. \textbf{92}, 256602 (2004).
\bibitem{JohnsonPRL2004} A. C. Johnson, C. M. Marcus, M. P. Hanson, and A. C. Gossard, Phys. Rev. Lett. \textbf{93}, 106803 (2004).

\bibitem{BulkaPRL2001} B. R. Bu\l ka and P. Stefa\'{n}ski, Phys. Rev. Lett. \textbf{86}, 225128 (2001).

\bibitem{TorioEPJB2004} M. E. Torio, K. Hallberg, S. Flach, A. E. Miroshnichenko, and M. Titov, Eur. Phys. J. B \textbf{37}, 399 (2004).

\bibitem{CSChuPRB1989} C. S. Chu and R. S. Sorbello, Phys. Rev. B \textbf{40}, 5941 (1989).

\bibitem{RuiZhuJPCM2013} R. Zhu, J. Phys.: Condens. Matter \textbf{25}, 036001 (2013).

\bibitem{BoykinPRB1992} T. B. Boykin, B. Pezeshki, and J. S. Harris, Phys. Rev. B \textbf{46}, 12769 (1992).

\bibitem{PorodAPL1992} W. Porod, Z. Shao, and C. S. Lent, Appl. Phys. Lett. \textbf{61}, 1350 (1992).

\bibitem{CardosoEPL2008} J. L. Cardoso and P. Pereyra, Europhys. Lett. \textbf{83}, 38001 (2008).

\bibitem{KobayashiPRB2004} K. Kobayashi, H. Aikawa, A. Sano, S. Katsumoto, and Y. Iye, Phys. Rev. B \textbf{70}, 035319 (2004).

\bibitem{LukyanchukNatMat2010} B. Luk¡¯yanchuk, N. I. Zheludev, S. A. maier, N. J. Halas, P. Nordlander, H. Giessen, and C. T. Chong, Nature Materials \textbf{9}, 707 (2010).

\bibitem{1ThoulessPRB1983} D. J. Thouless, Phys. Rev. B \textbf{27}, 6083 (1983).

\bibitem{4XiaoRMP2010}D. Xiao, M.-C. Chang, Q. Niu, Rev. Mod. Phys. \textbf{82}, 1959(2010).

\bibitem{3BrouwerPRB1998} P. W. Brouwer, Phys. Rev. B \textbf{58}, R10135 (1998).

\bibitem{2MoskaletsPRB2002} M. Moskalets, M.B\"{u}ttiker, Phys. Rev. B \textbf{66}, 035306 (2002).

\bibitem{38LiPRB1999} W. Li, L. E. Reichl, Phys. Rev. B\textbf{60}, 15732 (1999).

\bibitem{44Jiao-Hua DaiEPJB2014} J.-H. Dai and R. Zhu, Eur. Phys. J. B  \textbf{87}, 288 (2014).

\bibitem{EckardtRMP2017} A. Eckardt, Rev. Mod. Phys. \textbf{89}, 011004 (2017).

\bibitem{WeiYinDengNJP2015} W. Y. Deng, W. Luo, H. Geng, M. N. Chen, L. Sheng, and D. Y. Xing, New J. Phys. \textbf{17}, 103018 (2015).

\bibitem{UsajPRB2014} G. Usaj, P. M. Perez-Piskunow, L. E. F. F. Torres, and C. A. Balseiro, Phys. Rev. B \textbf{90}, 115423 (2014).

\bibitem{DelplacePRB2013} P. Delplace, \'{A}. G\'{o}mez-Le\'{o}n, and G. Platero, Phys. Rev. B \textbf{88}, 245422 (2013).

\bibitem{LopezPRB2012} A. L\'{o}pez, Z. Z. Sun, and J. Schliemann, Phys. Rev. B \textbf{85}, 205428 (2012).

\bibitem{16S.L.ZhuPRB2002} S. L. Zhu, Z.D. Wang, Phys. Rev. B\textbf{65}, 155313 (2002).

\bibitem{49RZhuJAP2015} R. Zhu, J.-H. Dai, Y. Guo, Journal of Applied Physics  \textbf{17},164306(2015).

\bibitem{26Y.-C. XiaoPLA2013} Y.-C. Xiao, W.-Y. Deng, W.-J. Deng, R. Zhu, R.-Q. Wang,Phys. Lett. A \textbf{377}, 817 (2013).

\bibitem{36San-JosePRB2011} P. San-Jose, E. Prada, S. Kohler, H. Schomerus, Phys.Rev. B\textbf{84}, 155408 (2011).

\bibitem{RuiZhuEPJB2016January} R. Zhu and M. Liu, Eur. Phys. J. B \textbf{89}, 2 (2016).

\bibitem{52S. FafardPRB1993} S. Fafard, Y. H. Zhang, and J. L. Merz,Phys. Rev. B\textbf{48}, 12308 (1993).

\bibitem{57KrausePRB1998} J. L. Krause, D. H. Reitze and G. D. Sanders,Alex V. Kuznetsov,Christopher J. Stanton,Phys. Rev. B \textbf{57} 9024 (1998).

\bibitem{55BetancourtRieraPhysica E2012} R. Betancourt-Riera, R. Betancourt-Riera, R. Riera,R. Rosas,Physica E \textbf{44}, 1152 (2012).

\bibitem{53S. M. SadeghiPRB2010} S. M. Sadeghi,W. Li,Phys. Rev. B\textbf{81}, 155317 (2010).

\bibitem{54SchedelbeckScience1997} G. Schedelbeck, W. Wegscheider, M. Bichler, G. Abstreiter, Science \textbf{278}, 1792 (1997).

\bibitem{45BlanterPR2000} Y. M. Blanter, M. B\"{u}ttiker, Phys. Rep.\textbf{336},1 (2000).

\bibitem{6MoskaletsPRB2004} M. Moskalets, M. B\"{u}ttiker, Phys. Rev. B \textbf{70}, 245305 (2004).

\bibitem{TworzydloPRL2006} J. Tworzyd{\l}o, B. Trauzettel, M. Titov, A. Rycerz, and C. W. J. Beenakker, Phys. Rev. Lett. {\bf 96}, 246802 (2006).

\bibitem{ZhuHuiPLA2017} R. Zhu and P. M. Hui, Phys. Lett. A \textbf{381}, 1971 (2017).

\bibitem{LutchynPRL2010} R. M. Lutchyn, J. D. Sau, and S. D. Sarma, Phys. Rev. Lett. \textbf{105}, 077001 (2010).


\bibitem{LiangFuPRL2008} L. Fu and C. L. Kane, Phys. Rev. Lett. \textbf{100}, 096407 (2008).

\bibitem{56Song HePRB1991}S. He, X. C. Xie, and S. Das Sarma,Phys. Rev. B \textbf{43} 9339 (1991).

\bibitem{HamamotoPRB2015} K. Hamamoto, M. Ezawa, and N. Nagaosa, Phys. Rev. B \textbf{92}, 115417 (2015).

\end{references}
\end{document}